\documentclass{article}

\usepackage{arxiv}

\usepackage[utf8]{inputenc} % allow utf-8 input
\usepackage[T1]{fontenc}    % use 8-bit T1 fonts
\usepackage{hyperref}       % hyperlinks
\usepackage{url}            % simple URL typesetting
\usepackage{booktabs}       % professional-quality tables
\usepackage{amsfonts}       % blackboard math symbols
\usepackage{nicefrac}       % compact symbols for 1/2, etc.
\usepackage{microtype}      % microtypography
\usepackage{lipsum}

\usepackage{amsmath}%, amssymb}
\usepackage{authblk}
\usepackage{comment}% \begin{comment} \end{comment}でコメントアウト
\usepackage{graphicx}
\usepackage{tocbibind}%しおりに目次と参考文献を生成
\usepackage[usenames]{color}
\newcommand{\en}[1]{{#1}}%範囲出す

\def\numberline#1{%(Appendix重なり防止)
   \setbox\z@\hbox{#1\ }%
   \ifdim\wd\z@<\@lnumwidth \hbox to\@lnumwidth{\unhbox\z@\hfil}%
   \else                    \box\z@
   \fi}
\makeatother
\title{Multicoding in neural information transfer suggested by mathematical analysis of the frequency-dependent synaptic plasticity in vivo}

\author[1, 3, 5]{Katsuhiko Hata 
\thanks {khata@kokushikan.ac.jp}
}
\author[2]{Osamu Araki}
\author[3]{Osamu Yokoi}
\author[1]{Tatsumi Kusakabe}
\author[4]{Yoshio Yamamoto}
\author[1]{Susumu Ito}
\author[5]{Tetsuro Nikuni}

\affil[1]{Department of Sports and Medical Science, Kokushikan University, Tokyo, Japan}
\affil[2]{Department of Applied Physics, Faculty of Science Division I, Tokyo University of Science, Tokyo, Japan}
\affil[3]{Department of Neuroscience, Research Center for Mathematical Medicine, Tokyo, Japan}
\affil[4]{Laboratory of Veterinary Biochemistry and Cell Biology, Faculty of Agriculture, Iwate University, Morioka, Japan}
\affil[5]{Department of Physics, Faculty of Science Division I, Tokyo University of Science, Tokyo, Japan}

\begin{document}

\maketitle

\begin{abstract}
Two elements of neural information processing have primarily been proposed: firing rate and spike timing of neurons. In the case of synaptic plasticity, although spike-timing-dependent plasticity (STDP) depending on presynaptic and postsynaptic spike times had been considered the most common rule, recent studies have shown the inhibitory nature of the brain in vivo for precise spike timing, which is key to the STDP. Thus, the importance of the firing frequency in synaptic plasticity in vivo has been recognized again. However, little is understood about how the frequency-dependent synaptic plasticity (FDP) is regulated in vivo. Here, we focused on the presynaptic input pattern, the intracellular calcium decay time constants, and the background synaptic activity, which vary depending on neuron types and the anatomical and physiological environment in the brain. By analyzing a calcium-based model, we found that the synaptic weight differs depending on these factors characteristic in vivo, even if neurons receive the same input rate. This finding suggests the involvement of multifaceted factors other than input frequency in FDP and even neural coding in vivo.
\end{abstract}

%%%%%%%%%%%%%%%%%%%%%%%%%%%%%%%%%%%%%%%%%%%%%%%%%%%%%%%%%%%%%%%%%%%%%%%%%%%%%%%%%%%%%%%%%%%%%%%%%%%%%%%%%%%%%%%%%%%%%%%%%%%%%%%%%%%%%%%%%%%%

\thispagestyle{empty}
%\newpage
%\tableofcontents%目次の生成
\newpage
\setcounter{section}{0}%数字{0}を{-1}とすれば、0章から開始される。
%%%%%%%%%%%%%%%%%%%%%%%%%%%%%%%%%%%%%%%%%%%%%%%%%%%%%%%%%%%%%%%%%%%%%%%%%%%%%%%%%%%%%%%%%%%%%%%%%%%%%%%%%%%%%%%%%%%%%%%%%%%%%%%%%%%%%%%%%%%%%%%%%%%%%%%%%%%%%%%%%%%%%%%%%%%%%%%%%%%%%%%%%%%%%%%%%%%%%%%%%%%%%%%%%%%%%%%%%%%%%%%%%%%%%%%%%%%%%%%%%%%%%%%%%%%%%%%%%%%%%%%%%%%%%%%%%%%%%%%%%%%%%%%%%%%%%%%%%%%%%%%%%%%%%%%%%%%%%%%%%%%%%%%%%%%%%%%%%%%%%%%%%%%%%%%%%%%%%%%%%%%%%%%%%%%%%%%%%%%%%%%%%%%%%%%%%%%%%%%%%%%%%%%%%%%%%%%%%%%%%%%%%%%%%%%%%%%%%%%%%%%%%%%%%%
%%%%%%%%%%%%%%%%%%%%%%%%%%%%%%%%%%%%%%%%%%%%%%%%%%%%%%%%%%%%%%%%%%%%%%%%%%%%%%%%%%%%%%%%%%%%%%%%%%%%%%%%%%%%%%%%%%%%%%%%%%%%%%%%%%%%%%%%%%%%%%%%%%%%%%%%%%%%%%%%%%%%%%%%%%%%%%%%%%%%%%%%%%%%%%%%%%%%%%%%%%%%%%%%%%%%%%%%%%%%%%%%%%%%%%%%%%%%%%%%%%%%%%%%%%%%%%%%%%%%%%%%%%%%%%%%%%%%%%%%%%%%%%%%%%%%%%%%%%%%%%%%%%%%%%%%%%%%%%%%%%%%%%%%%%%%%%%%%%%%%%%%%%%%%%%%%%%%%%%%%%%%%%%%%%%%%%%%%%%%%%%%%%%%%%%%%%%%%%%%%%%%%%%%%%%%%%%%%%%%%%%%%%%%%%%%%%%%%%%%%%%%%%%%%%%
\pagenumbering{arabic}
%%%%%%%%%%%%%%%%%%%%%%%%%%%%%%%%%%%%%%%%%%%%%%%%%%%%%%%%%%%%%%%%%%%%%%%%%%%%%%%%%%%%%%%%%%%%%%%%%%%%%%%%%%%%%%%%%%%%%%%%%%%%%%%%%%%%%%%%%%%%%%%%%%%%%%%%%%%%%%%%%%%%%%%%%%%%%%%%%%%%%%%%%%%%%%%%%%%%%%%%%%ここから本文%%%%%%%%%%%%%%%%%%%%%%%%%%%%%%%%%%%%%%%%%%%%%%%%%%%%%%%%%%%%%%%%%%%%%%%%%%%%%%%%%%%%%%%%%%%%%%%%%%%%%%%%%%%%%%%%%%%%%%%%%%%%%%%%%%%%%%%%%%%%%%%%%%%%%%%%%%%%%%%%%%%%%%%%%%%%%%%%%%%%%%%%%%%%%%%%%%%%%%%%%%%%%%%%%%%%%%%%%%%%%%%%%%%%%%%%%%%%

\section{Introduction}
\label{Introduction}

\en{
Synaptic plasticity in neural networks is a substrate of learning and memory, which includes both positive and negative components, i.e., both long-lasting enhancements and declines in the weight of synaptic transmission (long-term potentiation (LTP) and long-term depression (LTD)) \cite{RN55}. Many experimental studies have suggested two plausible mechanisms for the induction of the synaptic plasticity \cite{RN57, RN95}. The first is the frequency of spike trains, which has been studied in association with the Bienenstock, Cooper, and Munro (BCM) rule in classical research conducted approximately half a century ago \cite{RN95, RN18, RN58, RN59}. LTP is induced by high-frequency firing in presynaptic neurons, which produces large increases in postsynaptic calcium concentration \cite{RN58, RN59, RN60, RN61}. The low-frequency firing causes a modest increase in the calcium level, and thereby induces LTD \cite{RN62, RN63, RN64}. The second is the precise timing of presynaptic and postsynaptic firing, which has been investigated as spike-time-dependent plasticity (STDP) in numerous experimental and theoretical studies from approximately 20 years ago \cite{RN95, RN3, RN65}. LTP is induced by the presynaptic action potentials preceding postsynaptic spikes by no more than tens of milliseconds, whereas presynaptic firing that follows postsynaptic spikes produces LTD \cite{RN65, RN50, RN51, RN53, RN69, RN70}. The idea that STDP plays a central role in synaptic plasticity had been becoming mainstream.
}
\en{
Recent studies have reported, however, that in some cases, the environment in vivo may not be suitable for precise spike timing, which is key to the STDP. Pre- and post-synaptic neurons in the primary visual cortex and extrastriate cortex of awaking animals fire so irregularly that the timing of presynaptic and postsynaptic firing varies \cite{RN13, RN73, RN76}. Neurons and synapses in the cerebral cortex of rats receive a lot of background neuronal activity that is generated internally, which provides strong constraints on spike timing \cite{RN33, RN74, RN75}. In these environments, the firing rate, rather than the spike timing, is likely to be important for the synaptic plasticity and neural coding. For example, it has been demonstrated experimentally that the cerebral cortex in which there is a high level of internal noise uses a rate code \cite{RN75}, and it has been shown mathematically that synaptic changes are induced by variation of firing rate without any timing constraints \cite{RN13}. Moreover, firing variability, as well as the statistical properties of the spike frequency, may be important for real-time information processing \cite{RN44}. Based on these reports, the role of firing frequency in various aspects of neural information processing has again come into the limelight. Furthermore, in vivo characteristic factors such as the variation of the firing pattern, the difference of intracellular parameters, and internal noise have also been suggested to be important for synaptic plasticity and neural coding. However, how these factors are involved in the synaptic plasticity is poorly understood. In order to clarify this problem, we examined the role of the presynaptic input pattern, the intracellular calcium decay time constants, and the background synaptic activity in frequency-dependent synaptic plasticity (FDP) by analyzing a calcium-based model, which is one of the most compatible models with experimental results \cite{RN3}.
}

\en{
Currently, it is widely accepted that the calcium concentration in the postsynapse determines whether LTP or LTD is induced \cite{RN77, RN78, RN79, RN80}. A moderate elevation of intracellular calcium correlates with induction of LTD, whereas a larger increase correlates with LTP \cite{RN78, RN79}. Only if glutamate is released by presynaptic activity and if the postsynaptic membrane is depolarized sufficiently, calcium ions enter the cell through channels controlled by NMDA receptors \cite{RN3}. The depolarization of the postsynaptic membrane potential is due not only to excitatory postsynaptic potentials (EPSPs) generated by binding glutamate to the AMPA receptors but also to many kinds of background synaptic activities \cite{RN34, RN35, RN27, RN37}. These experimental events were formulated by Shouval et al. \cite{RN1} as a calcium-based model, which has been used in numerous studies.
}

\en{
In the present study, we investigated the FDP in vivo analytically and numerically using the calcium-based model. First, in order to investigate the FDP in neurons with in vivo-specific firing pattern, we used three types of firing, which are widely observed in the brain, that is, constant-inter-spike intervals (ISI) inputs, Poisson inputs, and gamma inputs. Next, the calcium decay time constant of in vivo neurons varies from cell to cell. Previous reports suggested that pyramidal neurons in superficial layers possess faster calcium dynamics than those in deep layers. Here, $\tau_{ca} \approx 40$ ms in layer II to IV neurons, whereas $\tau_{ca} \approx 100$ ms in layer V to VI neurons \cite{RN23, RN5}. In order to study the association of the calcium decay time constant with the FDP, we examined two kinds of neurons with time constants of 40 ms and 80 ms. Finally, neurons in vivo are constantly exposed to background synaptic activity \cite{RN34, RN37}. The frequency and magnitude of this activity vary depending on the location of the synapse and the level of neuronal activity \cite{RN34, RN37}. We therefore examined the correlation between the amplitude of background activity and the FDP. The findings in the present study may contribute to a detailed understanding of synaptic plasticity in in vivo brain.
}

%%%%%%%%%%%%%%%%%%%%%%%%%%%%%%%%%%%%%%%%%%%%%%%%%%%%%%%%%%%%%%%%%%%%%%%%%%%%%%%%%%%%%%%%%%%%%%%%%%%%%%%%%%%%%%%%%%%%%%%%%%%%%%%%%%%%%%%%%%%%%%%%%%%%%%%%%%%%%%%%%%%%%%%%%%%%%%%%%%%%%%%%%%%%%%%%%%%%%%%%%%ここから本文%%%%%%%%%%%%%%%%%%%%%%%%%%%%%%%%%%%%%%%%%%%%%%%%%%%%%%%%%%%%%%%%%%%%%%%%%%%%%%%%%%%%%%%%%%%%%%%%%%%%%%%%%%%%%%%%%%%%%%%%%%%%%%%%%%%%%%%%%%%%%%%%%%%%%%%%%%%%%%%%%%%%%%%%%%%%%%%%%%%%%%%%%%%%%%%%%%%%%%%%%%%%%%%%%%%%%%%%%%%%%%%%%%%%%%%%%%%%
%%%%%%%%%%%%%%%%%%%%%%%%%%%%%%%%%%%%%%%%%%%%%%%%%%%%%%%%%%%%%%%%%%%%%%%%%%%%%%%%%%%%%%%%%%%%%%%%%%%%%%%%%%%%%%%%%%%%%%%%%%%%%%%%%%%%%%%%%%%%%%%%%%%%%%%%
%%%%%%%%%%%%%%%%%%%%%%%%%%%%%%%%%%%%%%%%%%%%%%%%%%%%%%%%セクション区切り%%%%%%%%%%%%%%%%%%%%%%%%%%%%%%%%%%%%%%%%%%%%%%%%%%%%%%%%%%%%%%%%%%%%%%%%%%%%%%%%%%%%%%%%%%%%%%%%%%%%%%%%%%%%%%%%%%%%%%%%

\noindent
\newline
\newpage

\newpage

\section{Results}
\label{Results}

\subsection{
\en{Postsynaptic calcium concentration as a function of the presynaptic stimulation frequency with fixed interstimulus intervals
}
}

\begin{figure}[htbp]
 %\begin{center}
 \centering%\hspace{-25mm}
  \includegraphics[width=140mm,clip%, bb=0 0 926 655
  ]{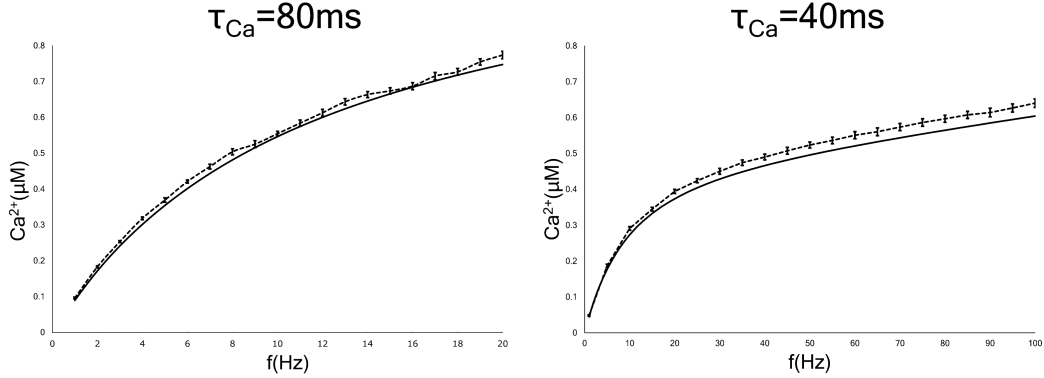}
 %\end{center}
  \vskip -\lastskip \vskip -10pt
 \caption{
\en{
Presynaptic firing rate-induced elevation of intracellular calcium concentration in two types of neurons with different time constants of calcium decay. The analytic solution is indicated by solid lines, while the results of numerical calculation are indicated by dotted lines. The calcium level increases more slowly in neurons with the short calcium decay time constant ($40$ ms) than in neurons with the long decay time constant ($80$ ms). This is also understood from Eq. (\ref{eq:<Ca_c(f)>}). Error bars indicate the standard error of the mean (SEM).
}
}
 \label{fig1}
\end{figure}

\en{
In Figure \ref{fig1}, we plot the analytical solution of $Ca$ in Eq. (\ref{eq:<Ca_c(f)>}) as a function of the input frequency $f$. We also plot the simulation results obtained by solving Eqs. (\ref{eq:W'(t)})-(\ref{eq:V_{bg}(t)}) numerically as a function of time and taking the time average of $Ca$ for each frequency. The analytical solution for the long-term behavior of calcium level agrees very well with the numerical simulation results. We adopted $\tau_{Ca}=80$ ms for a long calcium decay time constant and $\tau_{Ca}=40$ ms as a short calcium decay time constant. The calcium concentration as a function of input frequency increases slower for $\tau_{Ca}=40$ ms than for $\tau_{Ca}=80$ ms. Equation (\ref{eq:<Ca_c(f)>}) indicates that the calcium concentration at an arbitrary stimulation rate increases linearly for the calcium time constant $\tau_{Ca}$.
}

\subsection{
\en{
Approximate analytic solution of synaptic weight
}
}

\en{
Figure 2 shows the curve obtained by performing the integration in Eq. (\ref{eq:<W_{c}(f)>}). We also plot the results obtained by the numerical simulation, which agree qualitatively with the analytical results. These results suggest that the LTD/LTP threshold shifts to a lower frequency as the calcium time constant increases. Here, the LTD/LTP threshold is defined as the frequency at which the synaptic weight first returns to 1 after falling below 1 when the input frequency is increased from 0 Hz. This tendency can also be understood from Eq. (\ref{eq:<Ca_c(f)>}) as follows. Equation (\ref{eq:<Ca_c(f)>}) is written as $Ca(f)=\tau_{Ca} \cdot F(f)$, where $F(f)$ is a monotonically increasing function of $f$, so that $f$ can be formally expressed as $f=F^{-1}(Ca/\tau_{Ca})$. Equations (\ref{eq:Omega(ca)}) and (\ref{eq:approxW(f)}) indicate that when the synaptic strength is at the LTD/LTP threshold, the postsynaptic calcium level has a fixed value:
}

\begin{eqnarray}
Ca=\frac{1}{\beta}\log \frac{e^{\beta \alpha_1}-0.25e^{\beta \alpha_2}}{0.75} \ . \label{eq:Ca_th}
\end{eqnarray}

\en{
\noindent Substituting the numerical values of the parameters in Eq. (\ref{eq:Omega(ca)}) into Eq. (\ref{eq:Ca_th}), we obtained $Ca=0.54$ $\mu$M. Thus, the stimulation frequency when the synaptic weight reaches the LTD/LTP threshold is a monotonically increasing function of $1/\tau_{Ca}$.
}

\begin{figure}[htbp]
 %\begin{center}
 \centering%\hspace{-25mm}
  \includegraphics[width=140mm,clip%, bb=0 0 926 655
  ]{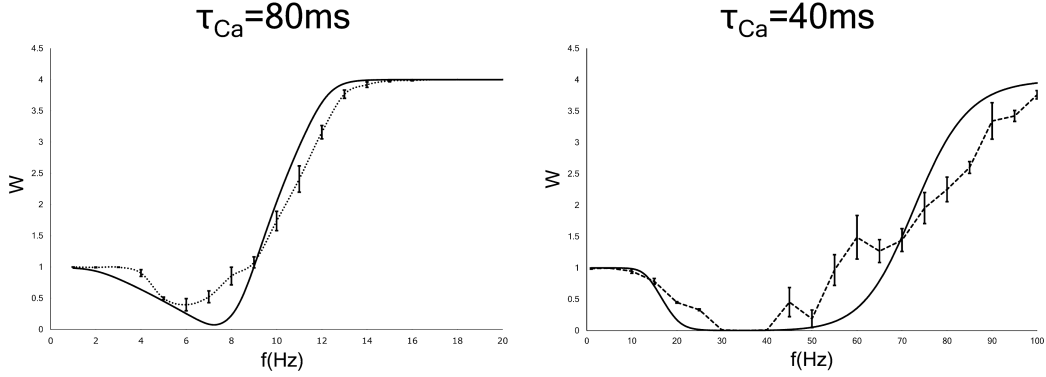}
 %\end{center}
  \vskip -\lastskip \vskip -10pt
 \caption{
 \en{
Synaptic strength in two types of neurons that have different calcium decay time constants as a function of the constant presynaptic stimulation frequency. The {\it x} axis indicates the input frequency, and the {\it y} axis represents normalized synaptic weights that are obtained after several hundreds of presynaptic spikes. The analytic solutions are indicated by solid lines, whereas the solutions provided by numerical calculation are indicated by dotted lines. Error bars indicate the SEM.
 }
  }
 \label{fig2}
\end{figure}

\subsection{
\en{
Postsynaptic calcium level and synaptic weight as functions of the average frequency of Poisson input
}
}
\en{
In several experimental studies on synaptic plasticity, the paradigms for inducing synaptic plasticity have consisted of constant-frequency stimulation trains, such as paired pulses or a tetanic stimulus. Neurons in vivo, however, are unlikely to experience such simple inputs. Rather, these neurons receive more complex input patterns in which ISIs are highly irregular \cite{RN19}. The most representative stimulation patterns that are not constant-frequency stimulation trains are the Poisson process and the gamma process. In fact, spike sequences similar to these processes are sometimes observed in neurons of brain \cite{RN3, RN20, RN21, RN22, RN30}. In this section, we discuss the results for the FDP of neurons with Poisson-distributed spike trains.
}

\begin{figure}[htbp]
 %\begin{center}
 \centering%\hspace{-25mm}
  \includegraphics[width=110mm,clip%, bb=0 0 926 655
  ]{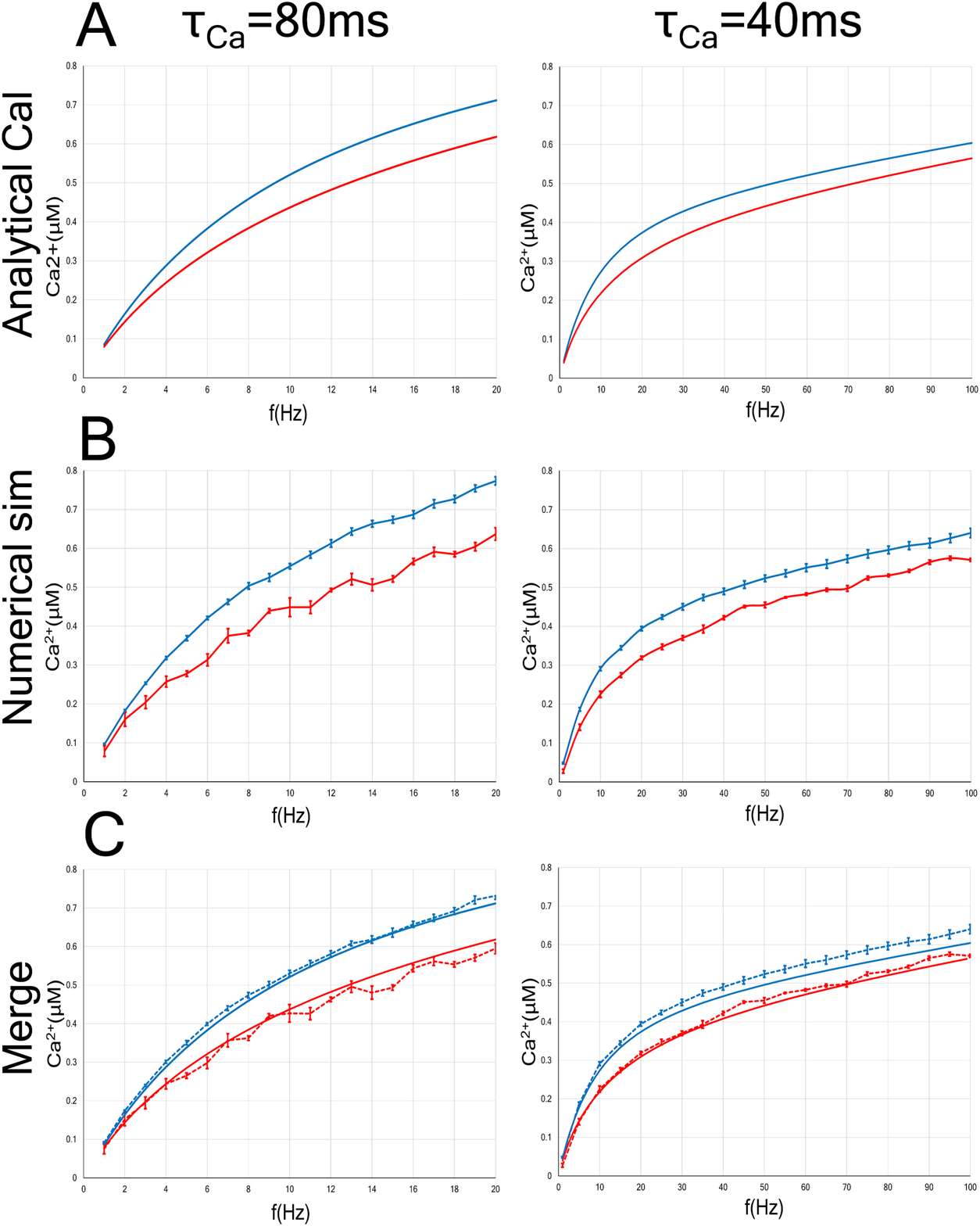}
 %\end{center}
 \vskip -\lastskip \vskip -10pt
 \caption{
\en{
Relation between the postsynaptic calcium concentration and the average frequency of presynaptic constant-ISI (or Poisson) input. As in Fig. \ref{fig1}, two types of neurons with different calcium time constants were examined. The analytic solution is shown in A and C, whereas the results of numerical simulation are shown in B and C. The blue (or red) lines indicate the calcium concentration of the postsynapse with constant-ISI (or Poisson) input. In the case of Poisson input, the increase in calcium concentration with respect to the average frequency is slower than in the case of constant-ISI input. This result is independent of the calcium time constant. Error bars indicate the SEM.
}
}
 \label{fig3}
\end{figure}

\en{
First, the calcium concentration at the postsynapse receiving Poisson input was calculated numerically, and is plotted in Fig. \ref{fig3}, in which the calcium concentration with constant-ISI input is also plotted for comparison. In the same manner, we examined two kinds of neurons with calcium time constants of 40 ms and 80 ms. The intracellular calcium concentration, regardless of the stimulation pattern, increases more gradually in the case of $\tau_{Ca}=40$ ms than in the case of $\tau_{Ca}=80$ ms. In addition, the calcium level with Poisson input increases more slowly than that with constant-ISI input, which is independent of the calcium time constant (Figs. \ref{fig3}B and \ref{fig3}C).
}

\begin{figure}[htbp]
 %\begin{center}
 \centering%\hspace{-25mm}
  \includegraphics[width=110mm,clip%, bb=0 0 926 655
  ]{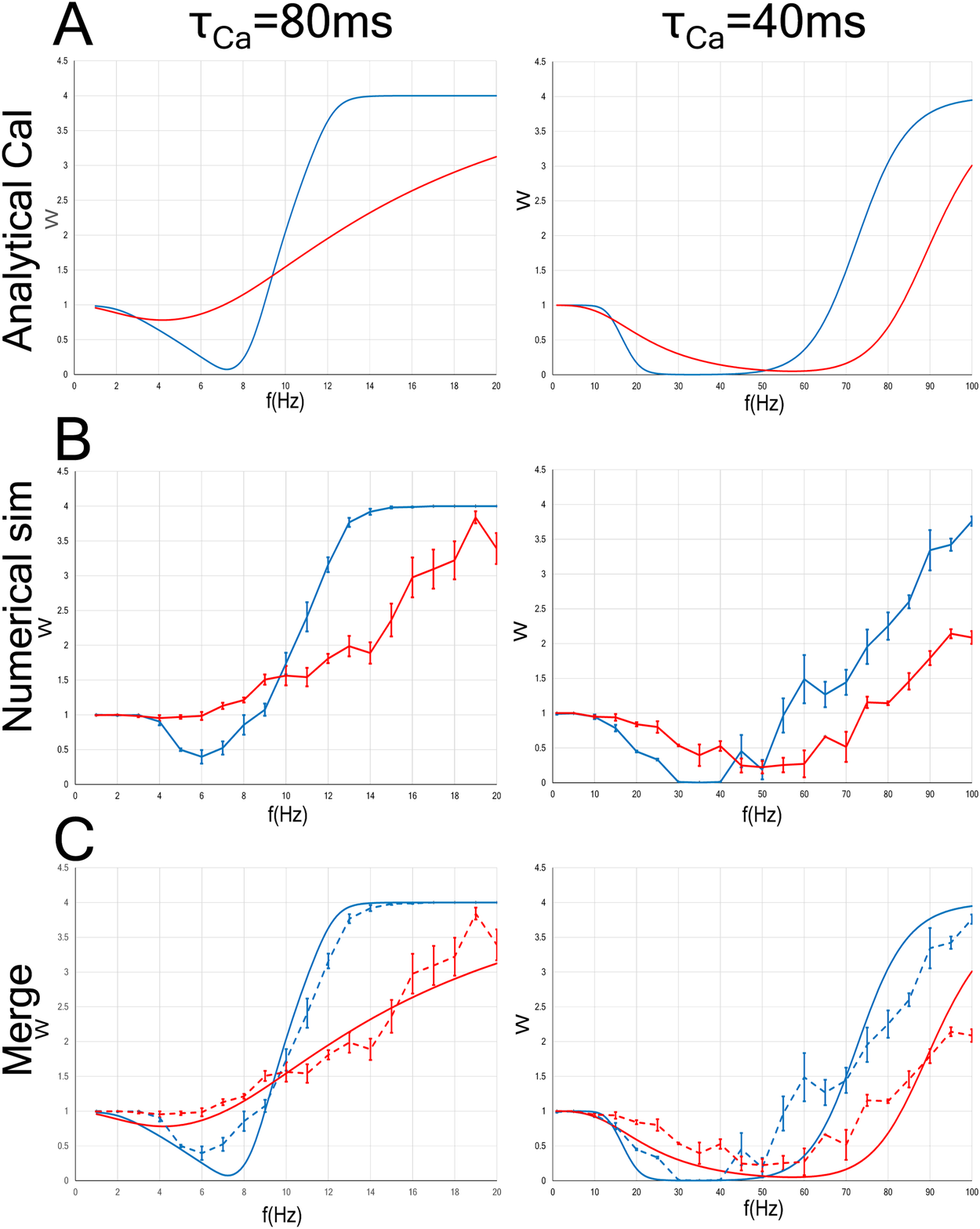}
 %\end{center}
 \vskip -\lastskip \vskip -10pt
 \caption{
 \en{
Synaptic strength for two types of neurons ($\tau Ca = 80$ ms and $\tau Ca = 40$ ms) as a function of the average rate of presynaptic stimulation. The {\it x} axis represents the input frequency, and the {\it y} axis represents normalized synaptic weights that are obtained after several hundreds of presynaptic spikes. The analytic solution is shown in A and C, whereas the results of numerical simulation are shown in B and C. Error bars indicate the SEM. The blue (or red) lines indicate the synaptic weights with constant-ISI (or Poisson) input. The synaptic weight with the Poisson input changes slowly compared to that with the constant-ISI input. As shown by the numerical simulation results, in neurons with $\tau Ca = 80$ ms, the Poisson input makes the LTD phase disappear, and only the LTP phase remains (B, left panel). On the other hand, in neurons with $\tau_{Ca} = 40$ ms, the LTD/LTP threshold moves to the right, and the LTD phase increases (B, right panel). These results are also qualitatively illustrated by analytical solutions (A). 
 }
 }
 \label{fig4}
\end{figure}

\en{
Next, we examined the strength of a synapse receiving Poisson input. In Fig. \ref{fig4}, we define the LTD phase or LTP phase as the range of frequency indicating LTD or LTP. When the calcium time constant is $80$ ms, interestingly, Poisson input makes the LTD phase disappear and the LTP phase is observed at any input frequency, whereas in the case of constant-ISI stimulation, the LTD phase still exists at roughly between 3 Hz and 9 Hz (see Fig. \ref{fig4}B, left panel). When the calcium time constant is 40 ms, unlike in the case of $\tau_{Ca}=80$ ms, changing the stimulus pattern from constant-ISI input to Poisson input shifted the LTD/LTP threshold to the right (see Fig. \ref{fig4}B, right panel). Since the firing rate observed in the brain is found to be at most approximately 112 Hz, we need only consider synaptic plasticity within 100 Hz \cite{RN31}. This consideration leads to the conclusion that Poisson input to a neuron with $\tau_{Ca}=40$ ms expands the LTD phase and narrows the LTP phase. These results can be well reproduced by approximate analytical solutions (Eqs. (\ref{eq:<Ca_{poi}(f)>}) and (\ref{eq:<W_{poi}(f)>})).
}

\begin{comment}
\addtocounter{figure}{-1}
%%% ↑次の figure 環境にも同じ図番号をつけるための細工．
%%% 分割してできた 2 個の figure 環境の一方でしか \caption を
%%% 用いない場合には，この細工は不要（figure カウンタは
%%% figure(*) 環境内の \caption の際に増加するので）．
\newpage

\begin{figure}[htbp]
\caption{
\en{
\noindent
(Continued) The {\it x} axis represents the input frequency, and the {\it y} axis represents normalized synaptic weights that are obtained after several hundreds of presynaptic spikes. The analytic solution is shown in A and C, whereas the results of numerical simulation are shown in B and C. Error bars indicate the SEM. The blue (or red) lines indicate the synaptic weights with constant-ISI (or Poisson) input. The synaptic weight with the Poisson input changes slowly compared to that with the constant-ISI input. As shown by the numerical simulation results, in neurons with $\tau Ca = 80$ ms, the Poisson input makes the LTD phase disappear, and only the LTP phase remains (B, left panel). On the other hand, in neurons with $\tau_{Ca} = 40$ ms, the LTD/LTP threshold moves to the right, and the LTD phase increases (B, right panel). These results are also qualitatively illustrated by analytical solutions (A).
}
}
\label{fig4}
\end{figure}
\end{comment}

\begin{figure}[htbp]
 %\begin{center}
 \centering%\hspace{-25mm}
  \includegraphics[width=140mm,clip%, bb=0 0 926 655
  ]{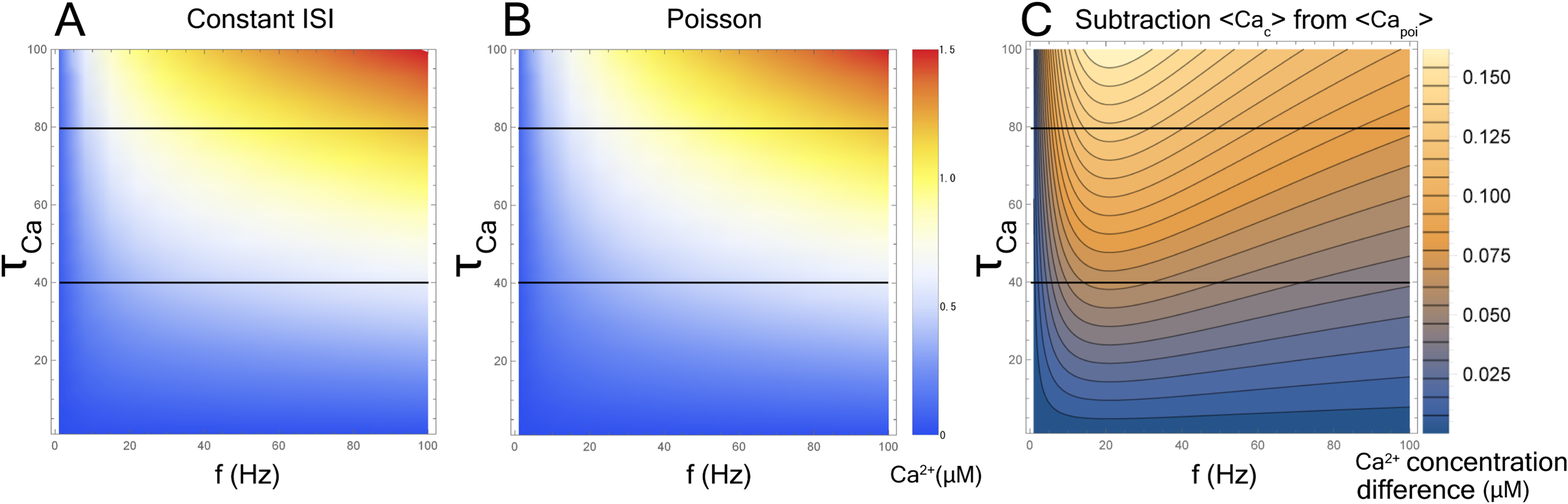}
 %\end{center}
 \vskip -\lastskip \vskip -10pt
 \caption{
\en{Two-dimensional density plot of post-synaptic calcium concentration as a function of $f$ and $\tau_{Ca}$. We illustrate the calcium concentration with the constant-ISI input expressed in Eq. (\ref{eq:<Ca_c(f)>}) (A)  and with the Poisson input expressed in Eq. (\ref{eq:<Ca_{poi}(f)>}) (B). (C) Density plot of $\left<Ca_c(f, \tau_{Ca}) \right>-\left<Ca_{poi}(f, \tau_{Ca}) \right>$. The horizontal lines on each figure suggest corresponding values at $\tau_{Ca}=40$ ms and $\tau_{Ca}=80$.
}
}
 \label{fig5}
\end{figure}

\en{
Analytical solutions for the calcium concentration with Poisson input (Eq. (\ref{eq:<Ca_{poi}(f)>})) are plotted in Fig. \ref{fig3}A. As shown in Fig. \ref{fig3}C, the solutions agree well with the numerical results and indicate that Poisson stimulation gently increases the calcium concentration, as compared to constant-ISI input. This property does not depend on the calcium decay time constant (Fig. \ref{fig5}).
}

\en{
Next, we obtained an approximate expression for the relation between the synaptic weight and the average stimulation rate. By assuming that the synaptic weight $W(t)$ converges to a stationary solution in the long-time scale (Eq. (\ref{eq:approxW(f)})), we obtain Eq. (\ref{eq:<W_{poi}(f)>}). Figures \ref{fig4}A and \ref{fig4}C show that the analytical expression agrees well with the results of the numerical simulation. Regardless of the $\tau_{Ca}$ value, the synaptic weight varies slowly by changing the stimulus pattern from constant-ISI input to Poisson input. This change in the stimulation pattern moves the LTD/LTP threshold to the left and narrows the LTD phase decrease for $\tau_{Ca}=80$ ms (Fig. \ref{fig4}A, left), whereas it has the opposite effect for $\tau_{Ca}=40$ ms (Fig. \ref{fig4}A, right).
}

\en{
Thus, the numerical and analytical studies indicate that the postsynaptic calcium concentration and synaptic strength receiving Poisson input behave differently from those receiving constant-ISI stimulation. At the same frequency, when $\tau_{Ca}=80$ ms, a synapse receiving Poisson input is more likely to be LTP than a synapse receiving constant-ISI input, and when $\tau_{Ca}=40$ ms, a synapse receiving Poisson input is more likely to be LTD. These findings suggest that the difference in input patterns (constant-ISI or Poisson input) and calcium decay time constant affects the output of FDP, i.e., LTD or LTP. In addition, this tendency to become LTP or LTD by changing the input pattern depends on the postsynaptic calcium decay time constant.
}

\subsection{
\en{
Postsynaptic calcium level and synaptic weight as a function of the average frequency of gamma process input
}
}

\en{
We studied the postsynaptic calcium concentration and synaptic load of neurons receiving gamma process inputs, which is one of the firing patterns observed in brain \cite{RN30, RN42}. Since the analytic solutions are qualitatively consistent with the simulation results so far presented in the present paper, we discuss the plasticity of synapses receiving gamma process input by only the analytic solutions. The postsynaptic calcium concentration of neurons that receive gamma process input is expressed by Eq. (\ref{eq:<Ca_{Gamma}(f)>}), where $\alpha$ is a shape parameter. The synaptic weight of the neurons receiving gamma process input is approximately expressed by Eq. (\ref{eq:<W_{Gamma}(f)>}) as a function of average input frequency.
}

\begin{figure}[htbp]
 %\begin{center}
 \centering%\hspace{-25mm}
  \includegraphics[width=140mm,clip%, bb=0 0 926 655
  ]{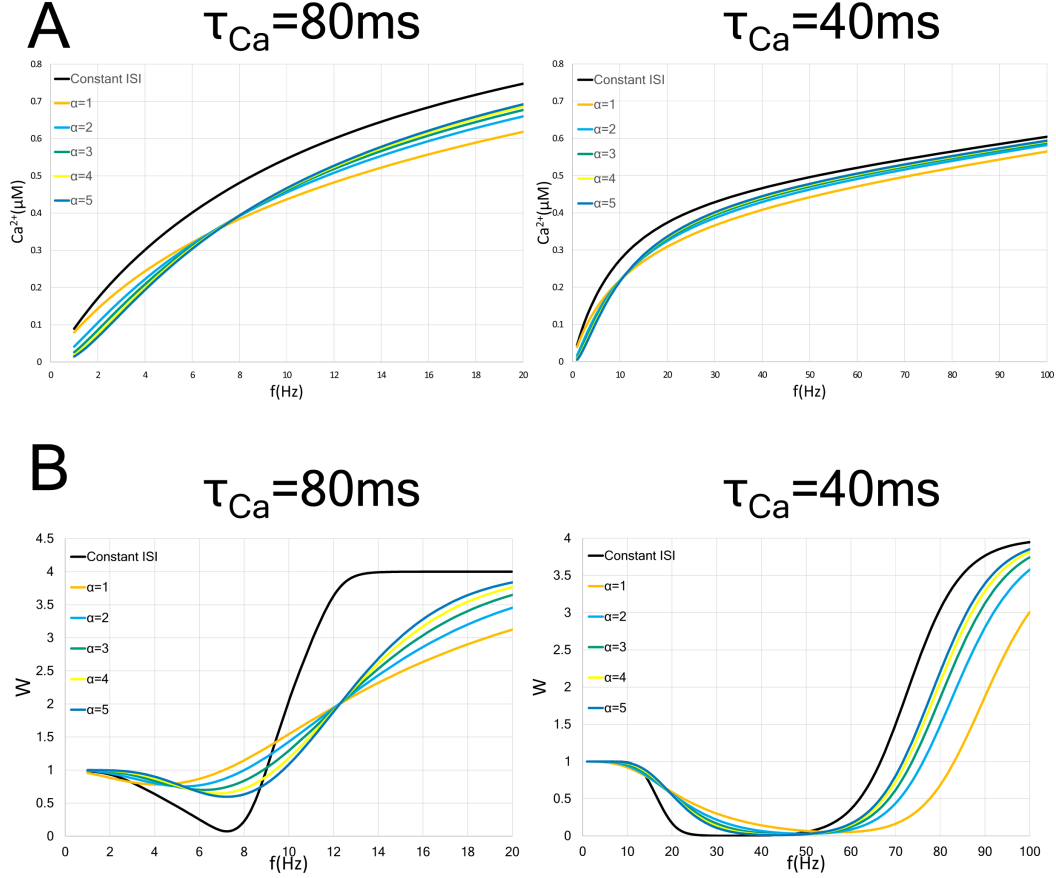}
 %\end{center}
 \vskip -\lastskip \vskip -10pt
 \caption{
\en{
Change in the postsynaptic calcium concentration and the weight in the synapse with gamma process input. We show two types of neurons with different time constants of calcium decay, $\tau_{Ca}=80$ ms and $40$ ms. In each graph, the black, orange, light blue, blue green, yellow, and blue lines indicate constant-ISI input, shape parameter $\alpha = 1$, $\alpha = 2$, $\alpha = 3$, $\alpha = 4$, and $\alpha = 5$, respectively.
(A) Relationship between the postsynaptic intracellular calcium concentration and input frequency $f$ in neurons stimulated with gamma process input. A graph of constant-ISI stimulation is shown as a control (black lines). The trace of the shape parameter $\alpha=1$ matches the graph of the Poisson input. As the value of the shape parameter increases, the calcium level increases is faster. 
(B) Approximate relationship between synaptic weight and mean input frequency in neurons with constant-ISI and gamma process inputs. The LTD/LTP threshold moves to a higher frequency in the case of $\tau_{Ca}=80$ ms and the moves lower in the case of $\tau_{Ca}=40$ ms as the value of the shape parameter becomes large.
}
}
 \label{fig6}
\end{figure}

\en{
This result for the calcium concentration is illustrated in Fig. \ref{fig6}A. As the shape parameter increases, the slope of the calcium concentration increases. The results for the synaptic weight are shown in Fig. \ref{fig6}B. When neurons with $\tau_{Ca}=80$ ms are stimulated by gamma process input, as the shape parameter $\alpha$ increases, the LTD/LTP threshold shifts to a higher frequency and the minimum value of the synaptic weight becomes smaller (Fig. \ref{fig6}B, left). When $\tau_{Ca}=40$ ms, the LTD/LTP threshold shifts to a lower frequency as the shape parameter $\alpha$ increases; on the other hand, the minimum value of the synaptic weight is approximately the same from $\alpha=1$ to $\alpha=5$ (Fig. \ref{fig6}B, right).
}

\en{
In summary, the postsynaptic calcium level with gamma process input increases slower than that with constant-ISI input, but increases faster than that with Poisson input. As the shape parameter increases, the increase in the calcium concentration becomes faster. The tendency to induce LTP or LTD by gamma process input depends on the shape parameter. These results suggest that the difference in input pattern as well as the shape parameter in gamma process input affects the synaptic weight.
}

\subsection{
\en{
Effect of increase in background synaptic activity receiving constant-ISI input
}
}

\en{
The postsynaptic terminals in neurons in vivo display intense background activity, which is characterized by fluctuations in the postsynaptic membrane potential. This background activity has at least three components: dendritic action potential, BPAPs, and voltage noise \cite{RN37, RN83}. The voltage noise includes the stochastic properties of ion channels, the random release of neurotransmitter, and thermal noise. The distance from the soma or the differences in the cortical layer, in which neurons are located, affects the frequency and size of the amplitude of the background synaptic activity \cite{RN34, RN35, RN27}.

In order to examine the FDP under various background synaptic activities, we first analytically and numerically calculated the dependence of the postsynaptic calcium concentration on the constant-ISI input under various frequencies of background Poisson input. The fluctuation of the membrane potential due to background synaptic activity is denoted by $V_{bg}$ in Eq. (\ref{eq:V_{bg}(t)}). Since $V_{bg}$ increases in proportion to the average frequency of the background synaptic activity $f_{bg}$, $H(V)$ in Eq. (\ref{eq:Ca(Delta t)}) is approximately expressed as a bivariate quadratic function of $f$ and $f_{bg}$. Thus, the postsynaptic calcium concentration is given as a function of $f$ and $f_{bg}$ as follows:
}

\begin{align}
\left<Ca_c(f, f_{bg}) \right> =& \tau _{Ca} f (\zeta_0+\zeta_1 f+\zeta_2 f_{bg}+\zeta_3 f^2+\zeta_4 f f_{bg}+\zeta_5 f_{bg}^2) \notag \\
& \times \sum_{j=f, s} I_j \tau_{j} \left[1-\exp \left(-\frac{1}{\tau_{j} \cdot f} \right) \right], \ \label{eq:<Ca(f, f_{bg})>}
\end{align}

\begin{figure}[htbp]
 %\begin{center}
 \centering%\hspace{-25mm}
  \includegraphics[width=140mm,clip%, bb=0 0 926 655
  ]{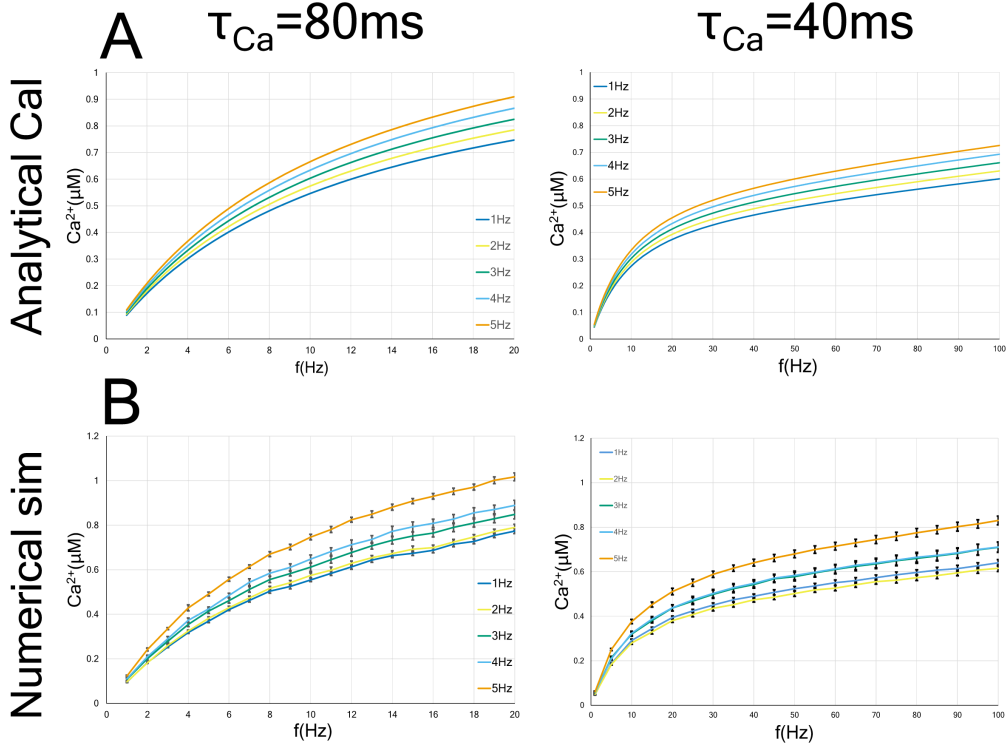}
 %\end{center}
 \vskip -\lastskip \vskip -10pt
 \caption{
 \en{
Postsynaptic calcium concentration in two types of neurons ($\tau Ca = 80$ ms and $\tau Ca = 40$ ms) as a function of the frequency of presynaptic input and of the background input. The ISI of the presynaptic input is constant. The background Poisson input with a frequency in the range of 1 to 5 Hz was applied. The analytic solution is shown in A, whereas the results of numerical simulation are shown in B. Error bars indicate the SEM.
 }
 }
 \label{fig7}
\end{figure}

\en{
\noindent where $\zeta_0=1.21 \times 10^{-2}$, $\zeta_1=2.97 \times 10^{-5}$, $\zeta_2=6.12 \times 10^{-4}$, $\zeta_3=3.52 \times 10^{-8}$, $\zeta_4=1.45 \times 10^{-6}$, and $\zeta_5=1.49 \times 10^{-5}$. Figure \ref{fig7}A plots Eq. (\ref{eq:<Ca(f, f_{bg})>}) using $\tau_{Ca}=80$ ms or $\tau_{Ca}=40$ ms. In both cases, the higher the average frequency of the background Poisson input is, the faster the rate of increase in the calcium concentration with synaptic input frequency becomes. As shown in Fig. \ref{fig7}B, qualitatively consistent results were obtained by numerical simulations.
}

\en{
We next analytically and numerically calculated the relation between the synaptic weight and the input frequency under various background input rates. The approximate analytic solution is obtained as follows:
}

\begin{align}
\left<W_{c}(f, f_{bg}) \right> = \int_{0}^{\infty} dx \int_{0}^{1} d\epsilon \ \delta(1-x) \Omega(Ca_{c}(f, f_{bg}, x, \epsilon| r_{Ca;c}, r_{j;c})). \label{eq:<W_{c}(f, f_{bg})>} 
\end{align}

\en{
\noindent Here, $Ca_{c}(f, f_{bg}, x, \epsilon| r_{Ca;c}, r_{j;c})$ is defined by Eq. (\ref{eq:<Ca(f, f_{bg})>}) in Eq. (\ref{eq:Ca(Delta t, x, epsilon| r_{Ca}, r_j)}). More explicitly, $Ca_{c}(f, f_{bg}, x, \epsilon| r_{Ca;c}, r_{j;c})$ is given by
}

\begin{align}
Ca_{c}(f, f_{bg}, x, \epsilon| r_{Ca;c}, r_{j;c}) =& (\zeta_0+\zeta_1 f+\zeta_2 f_{bg}+\zeta_3 f^2+\zeta_4 f f_{bg}+\zeta_5 f_{bg}^2) \notag \\
&
\sum_{j=f, s} I_j \tau_{0j} 
\left\{
\exp(-\frac{\epsilon}{\tau_j f})-\exp(-\frac{\epsilon}{\tau_{Ca} f})+\exp(-\frac{\epsilon}{\tau_{Ca} f}) \right. \notag \\
&\left. \left[\exp(-\frac{x}{\tau_{j} f})-\exp(-\frac{x}{\tau_{Ca} f}) \right] +
\exp(-\frac{x + \epsilon}{\tau_{Ca} f}) \left(\frac{r_j-r_{Ca}}{1-r_{Ca}} \right)
\right\}
. \label{eq:Ca_{c}(f, f_{bg}, x, epsilon| r_{Ca;c}, r_{j;c})}
\end{align}

\begin{figure}[htbp]
 %\begin{center}
 \centering%\hspace{-25mm}
  \includegraphics[width=140mm,clip%, bb=0 0 926 655
  ]{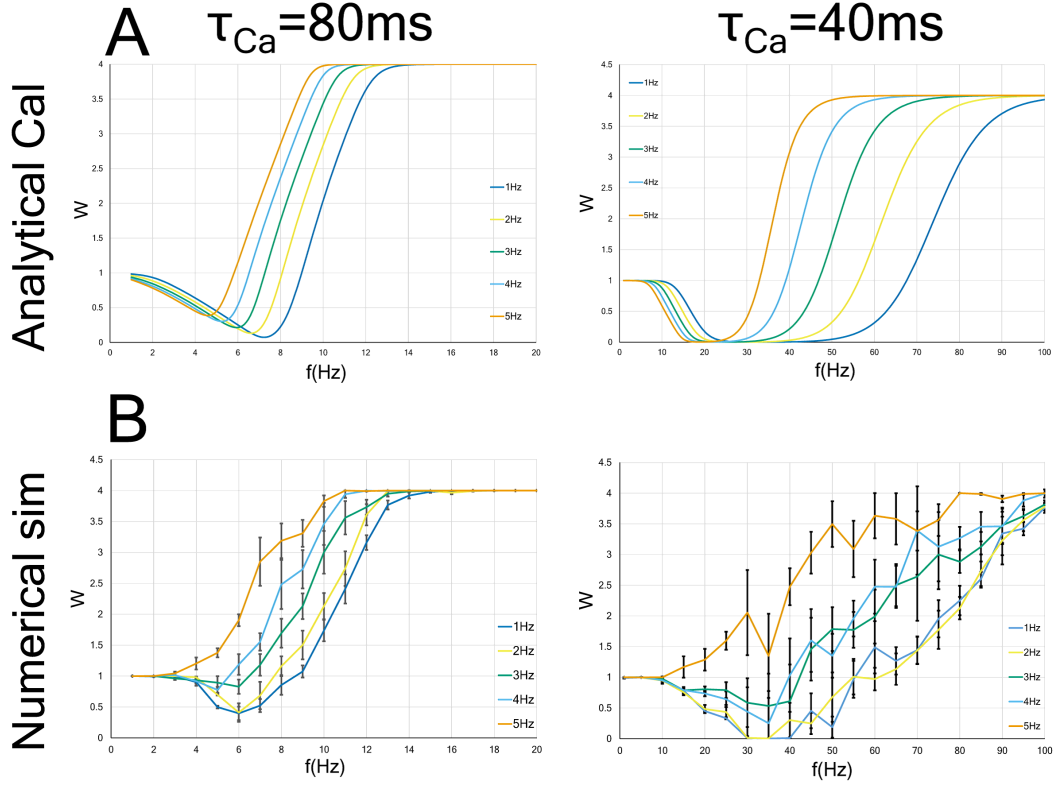}
 %\end{center}
 \vskip -\lastskip \vskip -10pt
 \caption{
 \en{
Synaptic strength as a function of the frequency of presynaptic constant-ISI input and of the background Poisson input, under the background Poisson input with a frequency in the range of 1 to 5 Hz. Two types of neurons ($\tau_{Ca} = 80$ ms and $\tau_{Ca} = 40$ ms) were examined. The analytic solution and the results of numerical calculation are shown in A and B, respectively. Error bars indicate the SEM.
 }
 }
 \label{fig8}
\end{figure}

\en{
The analytical solution (\ref{eq:<W_{c}(f, f_{bg})>}) is plotted in Fig. \ref{fig8}A, and the corresponding numerical solution is shown in Fig. \ref{fig8}B. Although two types of neurons with different calcium time constants were examined, the influence on the synaptic strengths by the increase of the background input level is qualitatively common to both types of neurons. In other words, the increase in the background input rate moves the LTD/LTP threshold to the left, decreases the LTD phase, and broadens the LTP phase.
}

\en{
Thus, upregulation of background synaptic activities leads to the enhancement of synaptic efficacy through the acceleration of the increasing rate of postsynaptic calcium concentration. These results suggest that the FDP output (LTP or LTD) varies depending on the magnitude of the applied background noise, even if the input frequency is the same.
}

%%%%%%%%%%%%%%%%%%%%%%%%%%%%%%%%%%%%%%%%%%%%%%%%%%%%%%%%%%%%%%%%%%%%%%%%%%%%%%%%%%%%%%%%%%%%%%%%%%%%%%%%%%%%%%%%%%%%%%%%%%%%%%%%%%%%%%%%%%%%%%%%%%%%%
%%%%%%%%%%%%%%%%%%%%%%%%%%%%%%%%%%%%%%%%%%%%%%%%%%%%%%%%%%%%%%%%%%%%%%%%%%%%%%%%%%%%%%%%%%%%%%%%%%%%%%%%%%%%%%%%%%%%%%%%%%%%%%%%%%%%%%%%%%%%%%%%%%%%%
%%%%%%%%%%%%%%%%%%%%%%%%%%%%%%%%%%%%%%%%%%%%%%%%%%%%%%%%%%%%%%%%%%%%%%%%%%%%%%%%%%%%%%%%%%%%%%%%%%%%%%%%%%%%%%%%%%%%%%%%%%%%%%%%%%%%%%%%%%%%%%%%%%%%%%%%%%%%%%%%%%%%%%%%%%%%%%%%%%%%%%%%%%%%%%%%%%%%%%%%%%%%%%%%%%%%%%%%%%%%%%%%%%%%%%%%%%%%%%%%%%%%%%%%%%%%%%%%%%%%%%%%%%%%%%%%%%%%%%%%%%%%%%%%%%%%%%%%%%%%%%%%%%%%%%%%%%%%%%%%%%%%%%%%%%%%%%%%%%%%%%%%%%%%%%%%%%%%%%%%%%%%%%%%%%%%%%%%%%%%%%%%

\noindent
\newline
\newpage
%\noindent
%\newline
%\newpage

\section{Discussion}
\label{Discussion}
\en{
Here, we summarize the findings of the present study:
(1) We obtained approximately analytical solutions of the intracellular calcium concentration and the synaptic weight as a function of the frequency of three kinds of input: constant-ISI, Poisson, and gamma process input. The latter two input patterns are often observed in vivo. 
(2) In all three input patterns, LTP occurs at a lower frequency as the calcium decay time constant increases. We used $80$ ms as the longer calcium decay time constant ($=\tau_{Ca}$) and $40$ ms as the shorter calcium decay time constant. 
(3) 
The intracellular calcium level increases more slowly in neurons with Poisson input than in neurons with constant-ISI input. At the same stimulation frequency, a synapse with a long calcium time constant tends to be strengthened (LTP) by changing the stimulation pattern from constant-ISI input to Poisson input, while a synapse with a short calcium time constant weakened (LTD).
(4) 
The calcium level with gamma process input increases faster than that with Poisson input but slower than that with constant-ISI input. Moreover, calcium level with gamma process also increases faster as the shape parameter grows. As the shape parameter increases, the LTD/LTP threshold moves to a higher frequency in $\tau_{Ca}=80$ ms neurons but moves to a lower frequency in $\tau_{Ca}=40$ ms neurons. The minimum value of the synaptic weight is smaller in $\tau_{Ca}=80$ ms neurons but is approximately constant in $\tau_{Ca}=40$-ms neurons as the shape parameter increases.
(5) 
The increase of background synaptic activities induces the acceleration of the increase rate of the calcium level and the enhancement of synaptic weight.

These findings indicate that the synaptic weight by FDP depends not only on input frequency but also on input pattern, shape parameter in gamma process input, calcium decay time constant, and background synaptic activity, which have been suggested to vary in vivo depending on the location, the internal state, and the external environment of the neuron \cite{RN37, RN23, RN5, RN30, RN42, RN83}. In the subsequent subsections, we discuss the involvement of these factors in synaptic plasticity and neural coding.
}

\en{
For a long time, there has been a debate on the nature of neural coding, which is primarily founded on the generation, propagation, and processing of spikes \cite{RN45, RN81, RN82}. The classical view of neural coding emphasizes the information carried by the rate at which neurons produce action potentials, whereas spike variability and background activity were ignored or treated as noise \cite{RN44, RN90, RN91}. In experimental and theoretical studies of recent decades, arguing the importance of the spike timing rather than the firing rate in neural coding, the spike variability and background activity are also considered as noise activities \cite{RN3, RN75}. However, the results of recent electrophysiological experiments on waking animals suggest that they are too large to be ignored for precise spike timing \cite{RN75,RN83}, leading to a renewed awareness of the importance of the rate coding, which is less affected by individual spike variability and background noise \cite{RN13, RN73}. Moreover, recent studies reveal the need for several simultaneous codes (multi-coding), including spike variability and fluctuation of membrane potential, as sources \cite{RN45, RN48, RN47, RN46}. Hence, the multi-coding hypothesis for the neural coding problem may be supported by the results of the present study, suggesting that not only firing rate but also firing variability, the internal parameters of neurons, and the magnitude of background synaptic activity could be important for neural coding and synaptic plasticity \cite{RN44, RN45}.
}

\en{
We found that the calcium decay time constant determines the plasticity outcome. In neurons with a long time constant, LTP is induced even by a small presynaptic rate (about 9 Hz), because the calcium concentration via the NMDA receptors increases faster in these neurons than in neurons with a short time constant (Figs. \ref{fig1} and \ref{fig2}). In neurons with a short time constant, LTP is not induced until the stimulation frequency is large (over about 50 Hz). This difference due to calcium dynamics is more pronounced when the stimulation pattern is set to Poisson or gamma process input (Figs. \ref{fig4} and \ref{fig6}).

The calcium decay time constant is closely related to the function of sodium-calcium exchangers (NCXs) \cite{RN85}. Sodium-calcium exchangers, which are expressed highly in dendrites and dendritic spines in a variety of brain regions \cite{RN84}, are controlled in activity by various intracellular and extracellular signaling molecules \cite{RN88} and are widely involved in many neural events from developmental processes to cognitive abilities \cite{RN86, RN87}. Thus, the calcium decay time constant differs depending on anatomical and physiological characteristics. Indeed, previous reports suggest that the calcium decay time constant varies with the depth of the cerebral cortex and that nitric oxide stimulates the increase of the calcium decay time constant in a cGMP-dependent manner \cite{RN5, RN88, RN89}. Our findings and those of previous studies suggest that, even with the same frequency, the synaptic plasticity induced thereby depends on the anatomical and physiological factors and that this difference becomes more prominent when the stimulation pattern is irregular.
}

\en{
Previous studies have demonstrated that applying an appropriate level of noise to the postsynapse results in the enhancement of the neural sensitivity and the improvement of signal detection in the central nervous system \cite{RN92, RN93}. Consistent with these findings, our research indicates that increased synaptic noise is more likely to induce LTP, regardless of the calcium time constant. Recently, the dendritic action potential has been considered as one of the main components of synaptic noise. In the record of the dendritic membrane potential of freely behaving rats, dendrite spikes accompanied by large subthreshold membrane potential fluctuations occur with high rates greater than the BPAP evoked in the soma \cite{RN83}. In addition, it has been shown in hippocampal synapses that even a single presynaptic burst induces LTP, provided dendritic action potentials are generated \cite{RN94}. These findings and our results indicate that inputs from other than the presynapse, such as background synaptic activity, including the BPAP and the dendritic action potential, are largely involved in synaptic plasticity, especially the generation of LTP. We cannot, however, conclude from our results that even a single presynaptic input induces LTP. It is necessary to conduct research in which single-burst-induced LTP is substantiated experimentally. Therefore, a mathematical model that further improves the model used in the present study should be constructed.
}

\en{
In conclusion, a problem regarding the FDP, namely, a firing rate abstraction, in which the temporal average of spikes is taken, is discussed, ignoring a large amount of extra information within the encoding window, such as the variation of firing pattern \cite{RN95, RN44, RN91}. This loss of information contrasts the encoding of rapidly changing neuronal activity observed in the brain \cite{RN95, RN44}. The present study showed theoretically that the output of synaptic plasticity in neurons receiving the same input frequency differs depending on the input pattern, the calcium time constant, and the background activity, which are related by neuron type and the anatomical and physiological condition in the brain. This finding suggests that information neglected in the view that only the firing rate induces the synaptic plasticity is also involved in the synaptic plasticity and neural coding. In the future, the ratio at which this information is related to synaptic plasticity and neural coding should be verified experimentally and theoretically.
}

\newpage
%%%%%%%%%%%%%%%%%%%%%%%%%%%%%%%%%%%%%%%%%%%%%%%%%%%%%%%%%%%%%%%%%%%%%%%%%%%%%%%%%%%%%%%%%%%%%%%%%%%%%%%%%%%%%%%%%%%%%%%%%%%%%%%%%%%%%%%%%%%%%%%%%%%%%%%%%%%%%%%%%%%%%%%%%%%%%%%%%%%%%%%%%%%%%%%%%%%%%%%%%%%%%%%%%%%%%%%%%%%%%%%%%%%%%%%%%%%%%%%%%%%%%%%%%%%%%%%%%%%%%%%%%%%%%%%%%%%%%%%%%%%%%%%%%%%%%%%%%%%%%%%%%%%%%%%%%%%%%%%%%%%%%%%%%%%%%%%%%%%%%%%%%%%%%%%%%%%%%%%%%%%%%%%%%%%%%%%%%%%%%%%%%%%%%%%%%%%%%%%%%%%%%%%%%%%%%%%%%%%%%%%%%%%%%%%%%%%%%%
%%%%%%%%%%%%%%%%%%%%%%%%%%%%%%%%%%%%%%%%%%%%%%%%%%%%%%サブセクション区切り%%%%%%%%%%%%%%%%%%%%%%%%%%%%%%%%%%%%%%%%%%%%%%%%%%%%%%%%%%%%%%%%%%%%%%%%%%%%%%%%%%%%%%%%%%%%%%%%%%%%%%%%%%%%%%%%%%%%%%
\section{Materials and methods}
\label{Method}
\subsection{Model}

\en{
We used a model for the FDP based on the calcium control hypothesis of Shouval et al., assuming that the change of the synaptic weight is fully determined by the postsynaptic calcium level \cite{RN1,RN2}. This model has been confirmed to integrate STDP observed in acute hippocampal slices within a single theoretical framework \cite{RN96}. Among the few studies that have analytically solved this hypothesis, Yeung et al. \cite{RN29} calculated the mean values of the calcium transients evoked by a spiking neuron. In the present study, we analytically derived the intracellular calcium concentration and synaptic weight with respect to the input frequency focusing only on the long-term behavior of the intracellular calcium concentration and synaptic weight.

We incorporated in vivo effects into the model as follows. First, in order to investigate the FDP in vivo, we focused on three types of firing pattern that are widely observed in the brain: constant-ISI (inter-spike intervals) inputs, Poisson inputs, and gamma inputs. Next, calcium decay time constant of in vivo neurons differs from cell to cell. Previous reports suggested that pyramidal neurons in superficial layers possess faster calcium dynamics than deep layers. In order to study the association of the calcium decay time constant with the FDP, we examined two kinds of neurons with time constants of 40 ms and 80 ms. Finally, in vivo neurons are always subjected to background activity. The frequency and magnitude of these neurons depend on the location of the synapse in the brain and the surrounding neuronal activity \cite{RN34, RN37}. Hence, we examined the correlation between the amplitude of background synaptic activity and the FDP.
}

\en{
The dynamics of the synaptic weight $W(t)$ are governed by
}
\begin{align}
\frac{d}{dt}W(t) = \eta(Ca(t))[\Omega(Ca(t))-W(t)] \ \ ,\label{eq:W'(t)}
\end{align}

\en{
\noindent where $Ca(t)$ represents the intracellular calcium concentration, and $\eta$ and $\Omega$ are functions of intracellular calcium concentration given by the following formulas:
}
\begin{align}
\eta(Ca) &= \left[\frac{p1}{p2+(Ca)^{p3}}+p4\right]^{-1} \ \ ,\label{eq:eta(ca)}\\
\Omega(Ca) &= 0.25+\mathrm{sig}(Ca-\alpha_2, \beta_2)-0.25 \mathrm{sig}(Ca-\alpha_1, \beta_1) \ \ \label{eq:Omega(ca)},
\end{align}

\en{
\noindent where
}
\begin{align}
 \quad \mathrm{sig}(x, \beta) = \exp(\beta x)/[1+\exp(\beta x)] \ \ \label{eq:sig},
\end{align}
\en{\noindent and we used the following parameters: $p1=0.1$ s, $p2=p1/10^{-4}$, $p3=3$, $p4=1$ s, $\alpha_1=0.35 \ \mu \mathrm{mol/dm^3}$, $\alpha_2=0.55 \ \mu \mathrm{mol/dm^3}$ and $\beta_1=\beta_2=80 \ \mu \mathrm{mol/dm^3}$ \cite{RN1,RN2}.

The dynamics of the intracellular calcium concentration are described as follows:
}
\begin{eqnarray}
\frac{d}{dt}Ca(t) &=& I_{\scalebox{0.5}{NMDA}}(t)-\frac{1}{\tau_{ca}}Ca(t) \label{eq:ca1} \ \ ,
\end{eqnarray}
\en{
\noindent where $\tau_{ca}$ is the calcium decay time constant. In order to investigate the relation between the calcium dynamics and the synaptic plasticity, we examined two kinds of neurons with time constants of 40 ms and 80 ms, which are known as representative values in pyramidal cells in the deep cortex (layers V to VI) and the superficial cortex (layers II to IV) \cite{RN23, RN5}.

In Eq. (\ref{eq:ca1}), $I_{\scalebox{0.5}{NMDA}}$ represents the calcium current via the NMDA receptor and is expressed as a function of time and postsynaptic potential as follows:
}
\begin{eqnarray}
I_{\scalebox{0.5}{NMDA}}(t, V) &=& H(V) \left[I_f \Theta(t) e^{(-t/\tau_f)} + I_s \Theta(t) e^{(-t/\tau_s)}\right] \label{eq:INMDA} \ \ .
\end{eqnarray}
\en{
\noindent Here, $\Theta(t)$ is the Heaviside step function and we choose the parameters $I_f=0.75$, $I_s=0.25$, $\tau_f=50$ ms, and $\tau_s=200$ ms, and $H(V)$ is given by
}
\begin{eqnarray}
H(V) &=& -P_0 \ G_{\scalebox{0.5}{NMDA}} \frac{(V-V_r)}{1+(Mg/3.57)\exp(-0.062 V)} \ \ , \label{eq:H(V)}
\end{eqnarray}
\en{
\noindent where we choose the parameters $P_0=0.5$, $G_{\scalebox{0.5}{NMDA}}=-1/140 \ \mathrm{\mu mol \cdot dm^{-3}/(m \cdot mV)}$, $Mg=3.57$, and a reversal potential for calcium ions of $Vr=130 \ \mathrm{mV}$ \cite{RN1}.
 Since $H(V)$ increases monotonically with the membrane potential $V$ before reaching a plateau at $V=27.1 \ \mathrm{mV}$, the higher the membrane potential the greater the calcium current through the NMDA receptor, $I_{\scalebox{0.5}{NMDA}}$, as long as $V < 27.1 \ \mathrm{mV}$.

The postsynaptic membrane potential is given as the sum of the resting membrane potential $V_{\rm rest}$, which is set to $-65$ mV, and the depolarization terms $V_{\rm epsp}+V_{\rm bg}$:
}
\begin{eqnarray}
V(t) &=& V_{\rm rest}+V_{\rm epsp}(t)+V_{\rm bg}(t) \ \ \label{eq:V(t)}.
\end{eqnarray}
\en{
The depolarization terms in Eq. (\ref{eq:V(t)}) include both EPSPs generated by binding glutamate to the AMPA receptors ($=V_{\rm epsp}$) and background contribution ($=V_{\rm bg}$), which describes the depolarization due to the factors other than EPSP. Here, $V_{\rm epsp}$ is expressed as
}
\begin{eqnarray}
V_{\rm epsp}(t) &=& \sum_{i} \Theta(t-t_i) \left[e^{-(t-t_i)/\tau_1}-e^{-(t-t_i)/\tau_2}\right] \ \ \label{eq:EPSP(t)},
\end{eqnarray}
\en{
\noindent where $t_i$ indicates the {\it i}-th presynaptic spike time, and the time constants are $\tau_1 = 50$ ms and $\tau_2 = 5$ ms \cite{RN1}. Here, $V_{\rm bg}$ is composed of the summation of the dendritic action potentials, the back propagating action potentials (BPAPs), and the voltage noise applied to the postsynapse. The amplitude of the depolarization generated at the postsynaptic dendritic spine by the BPAPs varies, decreasing exponentially with the distance from the soma, at which it is about $100$ mV relative to the synapse \cite{RN27, RN26}. The duration of the depolarization by BPAPs also differs among cell types \cite{RN28}.
Moreover, the noise level at dendritic spines has been reported to be similar to that measured at the soma \cite{RN25}. We took these previous studies into consideration in order to perform the numerical simulation and presumed that the spike trains by both BPAPs and voltage noise follow a homogeneous Poisson process. Thus, we simply expressed $V_{\rm bg}$ as follows:
}
\begin{eqnarray}
V_{\rm bg}(t) &=& s\sum_{k} \Theta(t-t_k) \left[e^{-(t-t_k)/\tau_1}-e^{-(t-t_k)/\tau_2}\right] \ \ \label{eq:V_{bg}(t)},
\end{eqnarray}
\en{
\noindent where $s = 20$ mV and $\{ t_k \}$ is a Poisson process with a frequency that varies depending on the simulation conditions. (In all simulations except for those of Figs. \ref{fig7} and \ref{fig8}, we used a Poisson process with a mean frequency of 1 Hz.)
}

\subsection{Numerical simulations}

\en{In the present study, we performed numerical simulations as well as analytical calculations in order to investigate the FDP. We used Wolfram Mathematica software in all simulations, and determined the dependence of both the calcium concentration and the synaptic weight on the stimulation frequency as follows. First, we repeatedly solved Eqs. (\ref{eq:W'(t)})-(\ref{eq:V_{bg}(t)}) numerically as a function of time for each frequency. The calcium concentration as a function of time obtained by this calculation is similar to the results of a previous paper \cite{RN29}. Next, after a period of $8.5 \times 10^4 $ ms, which is necessary for the system to reach a steady state, the average of the calcium level or the synaptic efficacy between $8.5 \times 10^4 $ ms to $9.0 \times 10^4 $ ms was calculated. When simulating with Poisson inputs, we performed the above calculations for at least three input patterns by changing the random seed, and took the average. The quantitative data are expressed as the mean of at least three independent experiments plus/minus the standard error of the mean (SEM).
}

\subsection{Derivation of the analytic solutions of the postsynaptic calcium concentration as functions of the average frequency of constant-ISI, Poisson, and gamma process inputs}

\en{
In order to investigate the dependence of the postsynaptic calcium concentration on the average presynaptic stimulation frequency of each input pattern, we developed an approximate analytical solution. By integrating Eq. (\ref{eq:ca1}), we can formally express the solution for $Ca(t)$ as
}

\begin{equation}
Ca(t)=\int_0^t e^{\frac{1}{\tau_{ca}}(s-t)}I_{\scalebox{0.5}{NMDA}}(s) ds \ . \label{eq:formal Ca(t)}
\end{equation}

\en{
\noindent Considering that the ion current through NMDAR ($I_{\scalebox{0.5}{NMDA}}$) is reset to zero each time presynaptic input is applied, Eq. (\ref{eq:INMDA}) is rewritten as follows for the interval between the presynaptic inputs $\hat{t}_k \leq s \leq \hat{t}_{k+1}$, where $\hat{t}_k$ is the time for $k$-th presynaptic input ($\hat{t}_0=0 \ \mathrm{ms}$):
}

\begin{align}
I_{\scalebox{0.5}{NMDA}}(s)=H(V)\left[I_f\Theta(s-\hat{t}_k) e^{-(s-\hat{t}_k)/\tau_f}+I_s\Theta(s-\hat{t}_k) e^{-(s-\hat{t}_k)/\tau_s}\right] \ . \label{eq:INMDA(s)}
\end{align}
\\
\\

\en{
Now, we make the following assumptions.
\\
\\
}

\en{
(Assumption 1) The time dependence of $H(V)$ can be neglected because it varies slowly in time compared to the other terms in Eq. (\ref{eq:INMDA(s)})\\
}

\en{
(Assumption 2) The spike interval fluctuates stochastically. If we define the average spike interval as $\Delta t$, $\hat{t}_k$ is written as follows:
\begin{align}
\hat{t}_k=\delta_k \Delta t + \hat{t}_{k-1}, \ \ \ \hat{t}_0=0.\label{eq:def:hat{t}_k}
\end{align}

\en{
\noindent Then,
}

\begin{align}
\hat{t}_k = \sum_{k'=1}^{k} \delta_{k'} \Delta t \ \ \ (k \geq 1). \label{eq:hat{t}_k}
\end{align}
}

\en{
\noindent Inserting Eq. (\ref{eq:INMDA(s)}) into Eq. (\ref{eq:formal Ca(t)}) with Assumption 1, we obtain
}

\begin{align}
Ca(t) =& H(V) \sum_{k=0}^{N} \int_{\hat{t}_k}^{\hat{t}_{k+1}} ds \left[ I_f\Theta(s-\hat{t}_k) e^{-(s-\hat{t}_k)/\tau_f}
\right. 
 \\
&+ \left. I_s\Theta(s-\hat{t}_k) e^{-(s-\hat{t}_k)\tau_s} \right] e^{\frac{1}{\tau_{ca}}(s-t)} \notag \\
=& H(V)[S_{N-1}^f(t)+T_{N}^f(t)+S_{N-1}^s(t)+T_{N}^s(t)] \ , \label{eq:Ca(t)}
\end{align}
\\
\en{
\noindent where we have separated the contributions from the {\it N}-th presynaptic input, $T_{N}^f(t)$ and $T_{N}^s(t)$ from the contributions from the first ${\it N}-1$ presynaptic inputs, $S_{N-1}^f(t)$ and $S_{N-1}^s(t)$:
}

\begin{align}
S_{N-1}^j(t) := \sum_{k=0}^{N-1} \int_{\hat{t}_k}^{\hat{t}_{k+1}} ds \ I_j \ e^{-\frac{1}{\tau_j}(s-\hat{t}_k)} \ e^{\frac{1}{\tau_{Ca}}(s-t)}, \ \ \ \ \ \ \mbox{($j=f$ or $s$)} \label{eq:S_{N-1}^j(t)} 
\end{align}

\en{
\noindent and
}

\begin{align}
T_{N}^j(t) := \int_{\hat{t}_N}^{t} ds \ I_j \ e^{-\frac{1}{\tau_j}(s-\hat{t}_N)} \ e^{\frac{1}{\tau_{Ca}}(s-t)}, \ \ \ \ \ \ \mbox{($j=f$ or $s$)} \ . \label{eq:T_{N}^j(t)}
\end{align}

\en{
\noindent Furthermore, we define $S_{k+1, k}^j(t)$ as
}

\begin{align}
S_{k+1, k}^j(t) :=& \int_{\hat{t}_k}^{\hat{t}_{k+1}} ds \ I_j \ e^{-\frac{1}{\tau_j}(s-\hat{t}_k)} \ e^{\frac{1}{\tau_{Ca}}(s-t)} \notag \\
=& I_j \tau_{0j} e^{-\frac{1}{\tau_{Ca}}(t-\hat{t}_k)} \left[e^{\frac{1}{\tau_{0j}}(\hat{t}_{k+1}-\hat{t}_{k})}-1 \right], \label{eq:S_{k+1, k}^j(t)}
\end{align}

\en{
\noindent where $\tau_{0f}$ and $\tau_{0s}$ are defined as follows:
}

\begin{align}
\frac{1}{\tau_{0j}} &:= \frac{1}{\tau_{Ca}}-\frac{1}{\tau_{j}}, \ \ \ \ \ \ \mbox{($j=f$ or $s$). } \label{eq:tau0j}
\end{align}

\en{
We write $t = \hat{t}_N + \epsilon \Delta t$, where $\epsilon \Delta t$ represents the time interval between the last spike time ($\hat{t}_N$) and the time to measure the calcium concentration ({\it t}). Substituting the formula into Eq. (\ref{eq:S_{k+1, k}^j(t)}), we obtain
}

\begin{align}
S_{k+1, k}^j(\hat{t}_N, \epsilon, \Delta t) :=& S_{k+1, k}^j(\hat{t}_N + \epsilon \Delta t) \notag \\
=& I_j \tau_{0j} e^{-\frac{1}{\tau_{Ca}}(\epsilon \Delta t + \hat{t}_N-\hat{t}_k)} \left(e^{\frac{1}{\tau_{0j}}\delta_{k+1} \Delta t}-1 \right) . \label{eq:S_{k+1, k}^j(hat{t}_N, epsilon, Delta t)}
\end{align}

\en{
\noindent In the case of $0\leq k \leq N-2$, we have
}

\begin{align}
S_{k+1, k}^j(\hat{t}_N, \epsilon, \Delta t) = I_j \tau_{0j} e^{-\frac{1}{\tau_{Ca}}\epsilon \Delta t} \left(e^{-\frac{1}{\tau_{j}}\delta_{k+1} \Delta t}-e^{-\frac{1}{\tau_{Ca}}\delta_{k+1} \Delta t} \right) \prod_{k'=k+2}^N e^{-\frac{1}{\tau_{Ca}}\delta_{k'} \Delta t}.  \label{eq:S_{k+1, k}^j(hat{t}_N + epsilon Delta t) k<N-2}
\end{align}

\en{
\noindent In the case of $k=N-1$, we have
}

\begin{align}
S_{N, N-1}^j(\hat{t}_N, \epsilon, \Delta t) = I_j \tau_{0j} e^{-\frac{1}{\tau_{Ca}}\epsilon \Delta t} \left(e^{-\frac{1}{\tau_{j}}\delta_{N} \Delta t}-e^{-\frac{1}{\tau_{Ca}}\delta_{N} \Delta t} \right).  \label{eq:S_{k+1, k}^j(hat{t}_N + epsilon Delta t) k=N-1}
\end{align}

\en{
Since we are interested in the long-term behavior of the calcium concentration and synaptic weights, but not in the fluctuations caused by each spike, we take the statistical average over one cycle.
Let $\delta_k$ in Assumption 2 obey the probability density function $\rho (\delta)$. Then the statistical averages of $e^{-\frac{1}{\tau_{Ca}}\delta_{k} \Delta t}$ and $e^{-\frac{1}{\tau_{j}}\delta_{k} \Delta t}$ can be written as
}

\begin{align}
r_{Ca} &:= \left<e^{-\frac{1}{\tau_{Ca}}\delta_{k} \Delta t} \right> = \int_0^\infty \rho (\delta) e^{-\frac{1}{\tau_{Ca}}\delta \Delta t} d \delta, \notag \\ r_{j} &:= \left<e^{-\frac{1}{\tau_{j}}\delta_{k} \Delta t} \right> = \int_0^\infty \rho (\delta) e^{-\frac{1}{\tau_{j}}\delta \Delta t} d \delta.   \label{eq:r_{Ca} r_{j}}
\end{align}

\en{
\noindent Hence, the statistical average of Eq. (\ref{eq:S_{k+1, k}^j(hat{t}_N + epsilon Delta t) k<N-2}) is given as
}

\begin{align}
\left<S_{k+1, k}^j(\hat{t}_N, \epsilon, \Delta t) \right>=I_j \tau_{0j} e^{-\frac{1}{\tau_{Ca}}\epsilon \Delta t} (r_j-r_{Ca}) r_{Ca}^{N-k-1}. \label{eq:<S_{k+1, k}^j>}
\end{align}

\en{
\noindent Summing from $k=0$ to $k=N-1$, the statistical average of Eq. (\ref{eq:S_{N-1}^j(t)}) is obtained as
}

\begin{align}
\left<S_{N-1}^j(\hat{t}_N, \epsilon, \Delta t) \right>=I_j \tau_{0j} e^{-\frac{1}{\tau_{Ca}}\epsilon \Delta t} (r_j-r_{Ca}) \frac{1-r_{Ca}^N}{1-r_{Ca}}. \label{eq:<S_{N-1}^j(hat{t}_N, epsilon, Delta t)>}
\end{align}

\en{
\noindent In order to obtain the long-term behavior of $\left<S_{N-1}^j \right>$, we take the limit $N \rightarrow \infty$. Since $r_{Ca}<1$, and thus $r_{Ca}^N \rightarrow 0$ as $N \rightarrow \infty$, we obtain
}

\begin{align}
\left<S_{N-1}^j(\epsilon, \Delta t) \right> \simeq I_j \tau_{0j} \frac{r_j-r_{Ca}}{1-r_{Ca}} \exp \left(-\frac{\epsilon \Delta t}{\tau_{Ca}} \right) \label{eq:<S_{N-1}^j(epsilon, Delta t)>}.
\end{align}

\en{
\noindent Similarly, using $t = \hat{t}_N + \epsilon \Delta t$ in Eq. (\ref{eq:T_{N}^j(t)}) and taking the statistical average, we obtain (in the limit $N \rightarrow \infty$)
}

\begin{align}
T_{N}^j(\epsilon, \Delta t) = I_j \tau_{0j} \left(e^{-\epsilon \Delta t/\tau_{j}}-e^{-\epsilon \Delta t/\tau_{Ca}} \right). \label{eq:T_{N}^j(epsilon, Delta t)}
\end{align}

\en{
\noindent Using Eqs. (\ref{eq:<S_{N-1}^j(epsilon, Delta t)>}) and (\ref{eq:T_{N}^j(epsilon, Delta t)}), we obtain the statistical average of the postsynaptic calcium concentration as
}

\begin{align}
\left<Ca(\Delta t, \epsilon) \right> = H(V) \sum_{j=f, s} I_j \tau_{0j} \left[\exp \left(-\frac{\epsilon \Delta t}{\tau_{j}} \right)-\frac{1-r_{j}}{1-r_{Ca}} \exp \left(-\frac{\epsilon \Delta t}{\tau_{Ca}} \right) \right] \ . \label{eq:Ca(Delta t,epsilon)}
\end{align}

\en{
\noindent Furthermore, the statistical average of this equation with respect to the observation time is given by
}

\begin{align}
\left<Ca(\Delta t) \right> = H(V) \sum_{j=f, s} I_j \tau_{0j} \left(r'_{j}-\frac{1-r_{j}}{1-r_{Ca}} r'_{Ca} \right), \label{eq:Ca(Delta t)}
\end{align}

\en{
\noindent where
}

\begin{align}
r'_{Ca}:=\left<e^{-\epsilon \Delta t/\tau_{Ca}} \right> \ \ \mbox{and} \ \ r'_{j}:=\left<e^{-\epsilon \Delta t/\tau_{j}} \right>. \label{eq:r'}
\end{align}

\subsubsection{Calcium concentration of constant-ISI input}

\en{
First, we calculate $r_{Ca}$, $r_{j}$, $r'_{Ca}$, and $r'_{j}$ for the constant-ISI input, which are denoted as $r_{Ca;c}$, $r_{j;c}$, $r'_{Ca;c}$, and $r'_{j;c}$, respectively. In this case, the probability density function is given by $\rho_{Ca;c}(x)=\rho_{j;c}(x)=\delta (1-x)$. Using this function in Eq. (\ref{eq:r_{Ca} r_{j}}), we obtain
}

\begin{align}
r_{Ca;c}=e^{-\frac{\Delta t}{\tau_{Ca}}}, \ \ \ r_{j;c}=e^{-\frac{\Delta t}{\tau_{j}}}. \label{eq:r const}
\end{align}

\en{
\noindent Since it is assumed that the sampling time follows a uniform distribution, $r'_{Ca}$ and $r'_{j}$ are expressed as follows:
}

\begin{align}
r'_{Ca;c}=\frac{\tau_{Ca}}{\Delta t} \left(1-e^{-\frac{\Delta t}{\tau_{Ca}}} \right), \ \ \ r'_{j;c}=\frac{\tau_{j}}{\Delta t} \left(1-e^{-\frac{\Delta t}{\tau_{j}}} \right). \label{eq:r' const}
\end{align}

\en{
\noindent Using Eqs. (\ref{eq:r const}) and (\ref{eq:r' const}), we obtain the statistical average of the postsynaptic calcium concentration as a function of the spike interval $\Delta t$ as follows:
}

\begin{align}
\left<Ca(\Delta t) \right> = H(V) \frac{\tau_{Ca}}{\Delta t} \sum_{j=f, s} I_j \tau_{j} \left[1-\exp \left(-\frac{\Delta t}{\tau_{j}} \right) \right]. \label{eq:<Ca(Delta t)>}
\end{align}

\en{
Note that $H(V)$ is a slowly changing and monotonically increasing function of the membrane potential in the vicinity of the resting membrane potential ($-65 {\rm mV}$), and the duration of depolarization by EPSP is approximately 50 to 100 ms at most. Therefore, the increase in the average membrane potential remains at approximately $5.4 {\rm mV}$, even in the case of the highest frequency, e.g., 100 Hz. The average membrane potential, moreover, increases linearly with the stimulation frequency. Thus, $H(V(\Delta t))$ is approximately expressed as a quadric function of $1/\Delta t(=f)$. With this approximation, we obtain the following expression:
}

\begin{align}
\left<Ca_c(f) \right> = \tau_{Ca} f (\gamma_0+\gamma_1 f+\gamma_2 f^2) \sum_{j=f, s} I_j \tau_{j} \left[1-\exp \left(-\frac{1}{\tau_{j} \cdot f} \right) \right]. \label{eq:<Ca_c(f)>}
\end{align}

\en{
\noindent Here, $\gamma_0= 1.28\times 10^{-2} {\rm mV}$, $\gamma_1= 3.20\times 10^{-2} {\rm mV \cdot ms}$, and $\gamma_2= 3.71\times 10^{-2} {\rm mV \cdot m^2}$. These values are determined by finding the relation between the input frequency and the time average of $V(t)$ in Eq. (\ref{eq:V(t)}) and by substituting the obtained values into the quadratic approximation of $H(V)$.
}

\subsubsection{Calcium concentration of Poisson input}

\en{
The time interval of the spike sequence according to the Poisson process follows an exponential distribution, the probability density function of which is given by $\rho_{Ca;poi}(x)=\rho_{j;poi}(x)=e^{-x}$. Then, we can calculate $r_{Ca}$ and $r_{j}$ for the Poisson input as
}

\begin{align}
r_{Ca;poi}=\frac{\tau_{Ca}}{\tau_{Ca}+\Delta t}, \ \ \ r_{j;poi}=\frac{\tau_{j}}{\tau_{j}+\Delta t}. \label{eq:r poi}
\end{align}

\en{
\noindent Since the spike interval fluctuates stochastically in the Poisson input, the observation time is considered to fluctuate with the same statistics. Then, $r'_{Ca}$ and $r'_{j}$ in the Poisson input, written as $r'_{Ca;poi}$ and $r'_{j;poi}$, are equal to $r_{Ca;poi}$ and $r_{j;poi}$, respectively. Substituting $r_{Ca;poi}$, $r_{j;poi}$, $r'_{Ca;poi}$, and $r'_{j;poi}$, we obtain the statistical average of the postsynaptic calcium concentration receiving Poisson input as a function of the average frequency as follows:
}

\begin{align}
\left<Ca_{poi}(f) \right> = \tau_{Ca} (\gamma_0+\gamma_1 f+\gamma_2 f^2) \sum_{j=f, s} I_j \frac{\tau_j f}{\tau_j f+1}. \label{eq:<Ca_{poi}(f)>}
\end{align}

\subsubsection{Calcium concentration of gamma process input}

\en{
The time interval of the spike sequence according to the gamma process follows a gamma distribution, the general formula for the probability density function of which is given as
}

\begin{align}
\rho_{Ca;\Gamma}(x; \alpha)=\rho_{j;\Gamma}(x; \alpha)=\frac{1}{\Gamma(\alpha)} x^{\alpha-1} e^{-x}, \label{eq:rho_{Gamma}}
\end{align}

\en{
\noindent where $\alpha$ is the shape parameter, and $\Gamma$ is the gamma function, which is given by
}

\begin{align}
\Gamma (\alpha):=\int_{0}^{\infty} t^{\alpha-1} e^{-t} dt.
\end{align}

\en{
\noindent Since, as in the Poisson input, the spike interval and the sampling time fluctuate with the same statistics, $r_{Ca}=r'_{Ca}=:r_{Ca:\Gamma}$ and $r_{j}=r'_{j}=:r_{j:\Gamma}$ in Eq. (\ref{eq:Ca(Delta t,epsilon)}). Thus, we obtain
}

\begin{align}
r_{Ca:\Gamma}=\left(\frac{\tau_{Ca}}{\tau_{Ca}+\Delta t} \right)^{\alpha}, \ \ \ r_{j:\Gamma}=\left(\frac{\tau_{j}}{\tau_{j}+\Delta t} \right)^{\alpha}. \label{eq:r Gamma}
\end{align}

\en{
\noindent Noting that the average spike interval of the gamma distribution input is $\alpha \Delta t$, we can express the statistical average of the postsynaptic calcium concentration with gamma process input as follows:
}

\begin{align}
\left<Ca_{\Gamma}(f) \right> = (\gamma_0+\gamma_1 f+\gamma_2 f^2) (\alpha f)^{\alpha} \sum_{j=f, s} I_j \tau_{0j} \left[
\frac{
\left(\frac{\tau_j}{\alpha \tau_j f+1} \right)^\alpha - \left(\frac{\tau_{Ca}}{\alpha \tau_{Ca} f+1} \right)^\alpha
}
{
1-\left(\frac{\alpha \tau_{Ca} f}{\alpha \tau_{Ca} f+1} \right)^\alpha
}
\right]. \label{eq:<Ca_{Gamma}(f)>}
\end{align}

\subsection{Derivation of the approximate analytic solutions for the synaptic weight as functions of the average frequency of constant-ISI, Poisson, and gamma process inputs}

\en{
According to the calcium control hypothesis reported by Shouval et al., the time derivative of the synaptic efficacy $W$ is expressed as a function of intracellular calcium concentration as indicated in Eqs. (\ref{eq:W'(t)})-(\ref{eq:Omega(ca)}) \cite{RN1}. Equation (\ref{eq:W'(t)}) indicates that the synaptic strength approaches an asymptotic value $\Omega(Ca(t))$ with time constant $1/\eta(Ca(t))$. The functional form of $\Omega(Ca(t))$ in Eq. (\ref{eq:Omega(ca)}) is based qualitatively on the notion that a moderate rise in calcium leads to a decrease in the synaptic weight, whereas a large rise leads to an increase in the synaptic weight. This notion is closely related to the BCM theory, which states that weak synaptic input activity results in a decrease in synaptic strength, whereas strong input leads to an increase in synaptic weight \cite{RN18, RN17}.\\
 
Although it is difficult to find the exact relation between the synaptic weight $W$ and the stimulation rate $f$ analytically, we can obtain an approximate relation by assuming that $W(t)$ converges to a stationary solution in the macroscopic time scale, i.e.,
}

\begin{eqnarray}
\lim_{t \to \infty} W(t, f) := \left< W(f) \right> \approx \left< \Omega(Ca(f)) \right> \ \ \label{eq:approxW(f)}.
\end{eqnarray}

\en{
\noindent In order to calculate $\left< \Omega(Ca(f)) \right>$, we express the postsynaptic calcium concentration as
}

\begin{align}
Ca(\Delta t, x, \epsilon| r_{Ca}, r_j) 
\approx &
\lim_{N \to \infty} H(V) \sum_{j=f, s} \left[T^j + S_{N, N-1}^j + \sum_{k=0}^{N-2} S_{k+1, k}^j \right] \notag \\
=& H(V) \sum_{j=f, s} I_j \tau_{0j} 
\left[
e^{-\frac{1}{\tau_j} \epsilon \Delta t}-e^{-\frac{1}{\tau_{Ca}} \epsilon \Delta t}
\right. \notag \\
&+e^{-\frac{1}{\tau_{Ca}} \epsilon \Delta t} \left(e^{-\frac{1}{\tau_{j}} x \Delta t}-e^{-\frac{1}{\tau_{Ca}} x \Delta t} \right)
\notag \\
&+
\left.
e^{-\frac{1}{\tau_{Ca}} (x + \epsilon) \Delta t} \left(\frac{r_j-r_{Ca}}{1-r_{Ca}} \right)
\right],
\label{eq:Ca(Delta t, x, epsilon| r_{Ca}, r_j)}
\end{align}

\en{
\noindent where $x = \delta_N$, $r_{Ca}$, and $r_j$ are defined in Eq. (\ref{eq:r_{Ca} r_{j}}). By substituting Eq. (\ref{eq:Ca(Delta t, x, epsilon| r_{Ca}, r_j)}) into the expression for $\Omega(Ca)$ in Eq. (\ref{eq:Omega(ca)}) and calculating the statistical average with respect to $x$ and $\epsilon$, we obtain an approximate analytical solution for the synaptic weight as a function of the average input frequency.
}

\en{
In the case of the constant-ISI input, the time interval of the spike sequence obeys the probability density function $\rho (x) = \delta(1-x)$. Moreover, the time interval from the last spike to the sampling time obeys a uniform distribution. Thus, we obtain the statistical average of the synaptic weight as a function of input frequency $f$ as follows:
}

\begin{align}
\left<W_{c}(f) \right> = \int_{0}^{\infty} dx \int_{0}^{1} d\epsilon \ \delta(1-x) \Omega(Ca(1/f, x, \epsilon| r_{Ca;c}, r_{j;c})). \label{eq:<W_{c}(f)>}
\end{align}

\en{
In the cases of the Poisson input and gamma process input, the spike interval as well as the time interval between the last spike and the observation time obey exponential and gamma distributions, respectively. Thus, the statistical average of the synaptic weight as a function of input frequency $f$ in these inputs are calculated as follows:
}

\begin{align}
\left<W_{poi}(f) \right> = \int_{0}^{\infty} dx \int_{0}^{\infty} d\epsilon \ e^{-(x + \epsilon)} \Omega(Ca(1/f, x, \epsilon| r_{Ca;poi}, r_{j;poi})), \label{eq:<W_{poi}(f)>}
\end{align}

\begin{align}
\left<W_{\Gamma}(f) \right> = \frac{1}{\Gamma(\alpha)^2} \int_{0}^{\infty} dx \int_{0}^{\infty} d\epsilon \ (x \epsilon)^{\alpha-1} e^{-(x+\epsilon)} \Omega(Ca(1/{\alpha f}, x, \epsilon| r_{Ca;\Gamma}, r_{j;\Gamma})). \label{eq:<W_{Gamma}(f)>}
\end{align}

\noindent
\newline
\newpage
%\newpage

\bibliographystyle{Science}
\bibliography{TMSpaper}

\end{document}